\documentclass[preprint]{elsarticle}
\usepackage{amsmath}
\usepackage{graphicx}
\usepackage{hyperref}
\usepackage{braket}
\usepackage{cancel}
\usepackage{color}

\begin{document}

\begin{frontmatter}

\title{Numerical modeling\\
  of exciton-polariton Bose--Einstein condensate\\
  in a microcavity}

\author[iftia]{Oksana~Voronych}

\author[iftia]{Adam~Buraczewski}

\author[ifpan]{Micha\l~Matuszewski}

\author[iftia,ifpan]{Magdalena~Stobi\'nska\corref{ms}}
\ead{magdalena.stobinska@gmail.com}

\cortext[ms]{Corresponding author}

\address[iftia]{Institute of Theoretical Physics and Astrophysics,
  University of Gda\'nsk,\\ ul.~Wita Stwosza 57, 80-952 Gda\'nsk,
  Poland}

\address[ifpan]{Institute of Physics, Polish Academy of Sciences,\\
  Al.~Lotnik\'ow 32/46, 02-668 Warsaw, Poland}

\begin{abstract}
A novel, optimized numerical method of modeling of an exciton-polariton superfluid in a semiconductor microcavity was proposed.  Exciton-polaritons are spin-carrying quasiparticles formed from photons strongly coupled to excitons.  They possess unique properties, interesting from the point of view of fundamental research as well as numerous potential applications.  However, their numerical modeling is challenging due to the structure of nonlinear differential equations describing their evolution. In this paper, we propose to solve the equations with a modified Runge--Kutta method of 4th order, further optimized for efficient computations. The algorithms were implemented in form of C++ programs fitted for parallel environments and utilizing vector instructions. The programs form the EPCGP suite which have been used for theoretical investigation of exciton-polaritons.
\end{abstract}

\begin{keyword}
exciton-polariton superfluid; Bose--Einstein condensate; microcavity; Gross--Pitaevskii equation; Runge--Kutta method
\end{keyword}

\end{frontmatter}

\section*{Program summary}

\begin{list}{}{\setlength{\leftmargin}{0pt}}

\item\emph{Program title:} EPCGP

\item\emph{Catalogue identifier:} \verb|EPCGP_v1_0|

\item\emph{Program summary URL:}
  \url{http://cpc.cs.qub.ac.uk/summaries/EPCGP_v1_0.html}

\item\emph{Program obtainable from:} CPC Program Library, Queen's
  University, Belfast, N. Ireland

\item\emph{Licensing provisions:} Standard CPC licence,
  \url{http://cpc.cs.qub.ac.uk/licence/licence.html}

\item\emph{No. of lines in distributed program, including test data,
    etc.:} 18748

\item\emph{No. of bytes in distributed program, including test data,
    etc.:} 200342

\item\emph{Distribution format:} ZIP

\item\emph{Programming language:} C++ with OpenMP extensions (main
  numerical program), Python (helper scripts)

\item\emph{Computer:} modern PC (tested on AMD and Intel processors),
  HP BL2x220

\item\emph{Operating system:} Unix/Linux and Windows

\item\emph{Has the code been vectorized or parallelized?:} yes
  (OpenMP)

\item\emph{RAM:} 200 MB for single run

\item\emph{Running time:} 6h for $100\kern.25em\mathrm{ps}$ evolution,
  depending on the values of parameters.

\item\emph{Classification:} 7. Condensed Matter and Surface Science; 7.7 Other Condensed Matter inc. Simulation of Liquids and Solids.

\item\emph{Nature of problem:} An exciton-polariton superfluid is a novel, interesting physical system allowing investigation of high temperature Bose--Einstein condensation of exciton-polaritons---quasiparticles carrying spin. They have brought a lot of attention due to their unique properties and potential applications in polariton-based optoelectronic integrated circuits. This is an out-of-equilibrium quantum system confined within a semiconductor microcavity. It is described by a set of nonlinear differential equations similar in spirit to the Gross--Pitaevskii (GP) equation, but their unique properties do not allow standard GP solving frameworks to be utilized. Finding an accurate and efficient numerical algorithm as well as development of optimized numerical software is necessary for effective theoretical investigation of exciton-polaritons.

\item\emph{Solution method:} A Runge--Kutta method of 4th order was employed to solve the set of differential equations describing exciton-polariton superfluids. The method was fitted for the exciton-polariton equations and further optimized. The C++ programs utilize OpenMP extensions and vector operations in order to fully utilize the computer hardware.

\end{list}

\section{Introduction}

\begin{figure}[ht]\centering
\includegraphics[height=5cm]{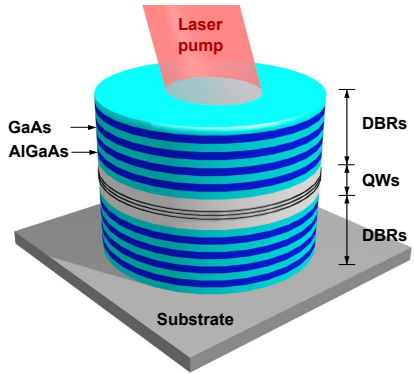}
\caption{Structure of a GaAs semiconductor microcavity.  Multiple
  layers of GaAs and AlGaAs form disributed Bragg reflectors (DBRs),
  which confine light inside quantum wells (QWs).}
\label{fig:microcavity}
\end{figure}

In this paper we propose a novel, optimized numerical method of modeling exciton-polariton superfluid in a semiconductor microcavity.  Excitons are electron-hole pairs, bound by the Coulomb force, behaving as a single electrically-neutral particle~\cite{Wannier1937,Mott1938}. Microcavities pumped by laser beams confine light in the form of a standing wave between highly reflective Bragg mirrors, which are made from multiple layers of different refractive index, see Fig.~\ref{fig:microcavity}. Between the reflectors there are located semiconductor quantum wells where the excitons are formed and can freely move in the X--Y plane. If the wells are located in the anti-nodes of the standing wave, they strongly couple to photons and thus, they form new quasiparticles---the exciton-polaritons.

\begin{figure}[ht]\centering
\includegraphics[height=3cm]{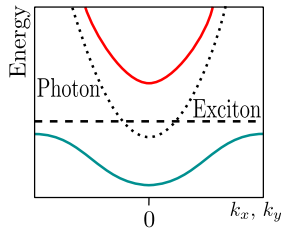}
\caption{Dispersion relation for exciton-polaritons in a semiconductor
  microcavity displaying upper (UP, red) and lower (LP, blue)
  polariton branches.}
\label{fig:dispersion}
\end{figure}

Fig.~\ref{fig:dispersion} depicts the dispersion relation for photons, excitons and exciton-polaritons in a microcavity.  It reveals two anti-crossing branches, called upper (UP) and lower polaritons (LP).  Exciton-polaritons are an out-of-equilibrium quantum system due to the interplay between their lifetime, up to $200\kern.25em\mathrm{ps}$, and laser pumping sustaining their number in the cavity. The compound nature of polaritons results in the fact that their effective mass is lower than the mass of a free electron, and in the regime of their low density they can be described as bosons with a spin degree of freedom~\cite{Hopfield1958,Agranovich1957,Kavokin2007}. Thus, in specific conditions, they form a quasi-particle counterpart of an atomic Bose-Einstein condensate (BEC)~\cite{Deng2010,Byrnes2014} and reveal superfluidity~\cite{Amo2009} in relatively high temperatures~\cite{Deveaud2007}.

Except for their amazing physical properties being a subject of the fundamental research, recently exciton-polaritons have brought a lot of attention due to their potential applications in optoelectronic integrated circuits, consisting of transistors~\cite{Liew2010}, spin-switches~\cite{Amo2010} and logic gates~\cite{Franson2007,Menon2010,Espinosa-Ortega2013}. Additionally, they can form localized nondiffracting X-waves~\cite{Voronych2016,Sedov2015} which could be used for transferring a classical signal between elements in the circuits.  Thus, polaritonics is regarded as a future of new photonic-electronic devices, which will be capable of processing information at a rate of terabits per second and frequencies in the range~$100\kern.25em\mathrm{GHz}$--$10\kern.25em\mathrm{THz}$~\cite{Feurer2007}.

The simplest physical model of the exciton-polariton superfluid is given by the Gross--Pitaevski equation (GPE). This is a nonlinear Schr\"odinger equation, which omits the quasi-particle nature of polaritons and which was primarily used for studying an akin discipline -- the physics of ultracold quantum bosonic gases (of atoms) and their BECs.
For this reason, over the years, a variety of numerical methods of solving GPEs were developed and implemented in software. They range from the most general, suitable for broad investigation of the gases, to specially fitted to specific systems and problems. Most papers devoted to numerical investigation of GPEs focused on their stationary solutions~\cite{Muruganandama2009}. Various condensate geometries~\cite{Chang2009}, simplifications and special cases~\cite{Chang2007} were taken into account.  Numerical methods involved finite-difference approach~\cite{Chang2009,Chang2008,Zeng2009}, bi-$k$-Lagrange elements~\cite{Li2009}, spectral collocation methods with Chebyshev
polynomials of the first and second kind~\cite{Jeng2013} as well as basis set expansion technique~\cite{Tiwari2006}.  Time-dependent equations were solved with implicit and semi-implicit Crank--Nicolson methods~\cite{Wang2010,Xu2012,Muruganandama2009,Madarassy2013,Mohammadi2014}, Euler scheme~\cite{Zeng2009}, third and fourth-order adaptive Runge--Kutta methods~\cite{Balac2013}, split-step finite difference method~\cite{Zeng2009} and time-splitting sine and Fourier pseudospectral methods~\cite{Wang2011,Vudragovic2012}.  In the latter case, space was discretized with second- and fourth-order finite differences, exponential splines~\cite{Mohammadi2014} or with Chebyshev--Tau spectral discretization method~\cite{Wang2010}.

As a result, several mature software packages were developed.  The OCTBEC utilizes optimal quantum control theory to model various BECs in Matlab~\cite{Hohenester2014}.  Similar libraries were prepared in Fortran~\cite{Muruganandama2009} and C programming languages~\cite{Vudragovic2012}.  The most advanced toolkit is the GPELab, implemented in Matlab~\cite{Antoine2014,Antoine2015}.  It combines various listed methods in order to solve both stationary and time-dependent GPEs and enables tackling sets of equations.  The hardware utilized for computations involved diverse platforms: OpenMP and MPI-based computer clusters~\cite{Satiric2016}, NVIDIA's CUDA parallel architecture~\cite{Dziubak2012,Loncar2016} as well as Sony PlayStation 3 Cell Broadband parallel systems~\cite{Edwards2009}.

Deeper insight into the physics of polaritons requires however taking into account their compound character and solving a GPE for a spinor polariton wave function, consisting of two independent components: the excitonic $\psi_x$ and photonic $\psi_c$ one. This turns the GPE into a system of two coupled equations of different kind, of which neither is a GPE itself and thus, methods developed for solving GPEs cannot be directly applied. Further including of the spin degree of freedom for polaritons results in the system of four equations. 

Here we present the EPCGP program suite which we have developed in order to support research on exciton-polaritons in semiconductor microcavities.  The suite utilizes our novel algorithm based on the Runge--Kutta method of fourth order, optimized for the equations describing exciton-polariton superfluid.  Moreover, program routines are able to gain from the parallel computing environment and vector operations, which significantly speeds up the computations.  It allows investigation of one- and two-dimensional systems.  We believe that use of EPCGP suite goes beyond the basic theoretical work and will also find applications in preparation of experiments and engineering of polaritonic circuits.

The paper is structured as follows.  Section~\ref{sec:theory} introduces the Reader to the equations describing the exciton-polariton superfluid.  Section~\ref{sec:numerics} goes into details of numerical computations, presenting the choice of algorithms, data structures and properties of the methods, such as their stability, computational complexity and error estimation.  Next, Section~\ref{sec:applications} presents a selection of interesting results obtained with our software.  Finally, Section~\ref{sec:suite} describes the actual suite code and goes through the process of preparation of input parameters, compilation and running the programs.

\section{Theoretical description of exciton-polariton superfluid}
\label{sec:theory}

\subsection{Polaritonic wave functions and the Gross--Pitaevskii equation}

Exciton-polariton superfluid is described by a composite wave function $\psi$, which consists of the photonic ($\psi_c$) and excitonic ($\psi_x$) parts~\cite{Sanvitto2012}.  In the spinor notation,
\begin{equation}
  \psi(\mathbf{x}, t) = \begin{pmatrix}
    \psi_c(\mathbf{x},t)\\
    \psi_x(\mathbf{x},t).
  \end{pmatrix}.
\end{equation}
$\psi_{c,x}$ are complex functions of space coordinate $\mathbf{x}$ and time $t$ such that $\lvert\psi_{c,x}(\mathbf{x}, t)\rvert^2$ is the distribution of quasiparticles in a space and $\int \lvert\psi_{c,x}(\mathbf{x}, t)\rvert^2\,d\mathbf{x}$ gives the number of quasiparticles in the system at given time instant. 

Formulation of equations describing the dynamics of polaritons requires solving a GPE, derived originally for an atomic BEC, for $\psi$
\begin{equation}
  i\hbar \dfrac{d\psi}{dt}
  =
  \left( -\dfrac{\hbar^2}{2m}\nabla^2 + V_{\text{ext}} +
  g\lvert\psi\rvert^2\right) \psi.
  \label{eq:basic_GP}
\end{equation}
Here $m$ is the mass of quasiparticles or atoms in the condensate, $V_{\text{ext}}$ is an external potential and $g$ quantifies strength of nonlinear interactions. Symbols $\hbar$, $i$ and $\nabla^2$ denote the reduced Planck constant, imaginary unit and the nabla operator $\nabla^2=\sum_{i=1}^n\tfrac{\partial^2}{\partial x_i^2}$, respectively.  

It is worth remembering that the GPE is a semi-classical equation, derived under assumption that a light beam pumping a microcavity is classical, and it describes correctly the exciton-polaritons in the regime of their low density, where to a good approximation polaritons behave as bosons.  There are two kinds of solutions of the GPE: the stationary, which describes the state of minimized energy, and the time-dependent, which allows to observe the dynamics of the system.  Their derivations are included in the graduate courses of physics and there is a lot of literature devoted to this topic~\cite{Rogel-Salazar2013}. Since we are interested in the evolution of the exciton-polarion superfluid, from now on we will focus solely on the time-dependent solutions.

In case of a semiconductor microcavity pumped by a laser pump 
\begin{equation}
F(\mathbf{x},t)=F_p\,e^{i\,(\mathbf{k}_p\cdot\mathbf{x}-\omega_p\,t)}\,e^{-\frac{(\mathbf{x}-\mathbf{x}_0)^2}{2w_x^2}},
\end{equation}
where $F_p$ is the field amplitude, $\mathbf{k}_p$ its momentum (plane profile), $\omega_p$ is the frequency, $\mathbf{x}_0$ is the coordinate of center of Gaussian laser spot on the sample and $w_x$ is its spread, Eq.~(\ref{eq:basic_GP}) takes the following matrix form~\cite{Kavokin2007}
\begin{multline}
  \label{eq:gp_matrix1}
  i\hbar\,\dfrac{d}{dt}\begin{pmatrix}
    \psi_c(\mathbf{x},t)\\
    \psi_x(\mathbf{x},t)
  \end{pmatrix}
  =
  \begin{pmatrix}
    F(\mathbf{x},t)\\
    0
  \end{pmatrix}
  +{}\\
  {}+\left[
    h^0+\begin{pmatrix}
      V_c(\mathbf{x})-i\hbar\,\frac{\gamma_c}{2}& 0\\
      0& V_x(\mathbf{x})-i\hbar\,\frac{\gamma_x}{2}
      +g\,\lvert\psi_x(\mathbf{x},t)\rvert^2
    \end{pmatrix}
  \right]
  \begin{pmatrix}
    \psi_c(\mathbf{x},t)\\
    \psi_x(\mathbf{x},t)
  \end{pmatrix}.
\end{multline}
The microcavity is characterized by the following parameters: $\gamma_c$ and $\gamma_x$--- the decay rates (loss rates) for photons and excitons, $g$---the strength of nonlinear exciton interaction, $V_c(\mathbf{x})$, $V_x(\mathbf{x})$---the single particle potentials acting on photons and excitons, and $\Omega_R$---the Rabi frequency. The single-particle Hamiltonian $h^0$ is given by
\begin{align}
  h^0={}&\begin{pmatrix}
    \omega_c(-i\nabla)& \Omega_R\\
    \Omega_R& \omega_x(-i\nabla)
  \end{pmatrix},\\
  \omega_c(-i\nabla)={}&\omega_c^0-\frac{\hbar^2\nabla^2}{2m_c},\\
  \omega_x(-i\nabla)={}&\omega_x^0-\frac{\hbar^2\nabla^2}{2m_x},
\end{align}
where $\omega_c(-i\nabla)$ is the cavity mode energy dispersion, $\omega_c^0$ is the cavity mode energy, $m_c$ is the effective mass of a polariton (usually of the order of $m_c=10^{-5}\cdot m_0$, $m_0$ being the mass of a free electron), $\omega_x(-i\nabla)$ is the exciton dispersion, $\omega_x^0$ is the exciton energy and $m_x$ is the effective mass of an exciton.

Although Eq.~(\ref{eq:gp_matrix1}) fully describes the evolution of exciton-polaritons, it includes a number of parameters which are usually unnecessary for investigation of the system in practice.  For example, the single particle potentials $V_c(\mathbf{x})$, $V_x(\mathbf{x})$ may be neglected in some situations and the exciton mass $m_x$ is regarded as infinite compared to the mass of a polariton.  We can also replace $\omega_p$, $\omega_c^0$ and $\omega_x^0$ with two parameters representing detuning of the pump field $\delta_{\omega}$ and detuning of polaritons $\delta$ from the cavity mode frequency $\omega_c^0$. Additionally, we assume that $\mathbf{x}=0$ lies in the center of the laser spot $\mathbf{x}_0$.  Summarizing, in our further discussion we take
\begin{align*}
  V_x=V_c={}&0,\\
  \omega_p\to{}&\omega_p+\omega_c^0=\delta_{\omega},\\
  \omega_c^0\to{}&\omega_c^0-\omega_c^0=0,\\
  \omega_x^0\to{}&\omega_x^0-\omega_c^0=\delta,\\
  \omega_c(-i\nabla)={}&-\frac{\hbar^2\nabla^2}{2m_c},\\
  \omega_x(-i\nabla)={}&\delta,\\
  \mathbf{x}_0={}&0.
\end{align*}
Under these assumptions Eq.~(\ref{eq:gp_matrix1}) is simplified to
\begin{multline}
  \label{eq:gp_matrix2}
  i\hbar\,\frac{d}{d t}\begin{pmatrix}
    \psi_c(\mathbf{x},t)\\
    \psi_x(\mathbf{x},t)
  \end{pmatrix}
  =
  \begin{pmatrix}
    F(\mathbf{x},t)\\
    0
  \end{pmatrix}
  +{}\\
  {}+\left[
    h^0+\begin{pmatrix}
      -i\hbar\,\frac{\gamma_c}{2}& 0\\
      0& -i\hbar\,\frac{\gamma_x}{2}
      +g\,\lvert\psi_x(\mathbf{x},t)\rvert^2
    \end{pmatrix}
  \right]\begin{pmatrix}
    \psi_c(\mathbf{x},t)\\
    \psi_x(\mathbf{x},t)
  \end{pmatrix},
\end{multline}
where
\begin{align}
  F(\mathbf{x},t)={}&F_p\,e^{i\,(\mathbf{k}_p\cdot\mathbf{x}-\delta_{\omega}\,t)}\,e^{-\frac{\mathbf{x}^2}{2w_x^2}},\\
  h^0={}&\begin{pmatrix}
    -\tfrac{\hbar^2\nabla^2}{2m_c}& \Omega_R\\
    \Omega_R& \delta
  \end{pmatrix}.
\end{align}
Next, we rewrite Eq.~(\ref{eq:gp_matrix2}) into a more convenient form 
\begin{align}
  \label{eq:gp_linear_psic1}
  i\hbar\,\frac{d}{dt}\,\psi_c(\mathbf{x},t)={}&
  F(\mathbf{x}, t)+\Omega_R\,\psi_x(\mathbf{x},t)
  +\left(-i\hbar\,\frac{\gamma_c}{2}-\frac{\hbar^2\nabla^2}{2m_c}\right)\,
  \psi_c(\mathbf{x},t),\\
  \label{eq:gp_linear_psix1}
  i\hbar\,\frac{d}{dt}\,\psi_x(\mathbf{x},t)={}&
  \Omega_R\,\psi_c(\mathbf{x},t)
  +\left(-i\hbar\,\frac{\gamma_x}{2}+g\,\lvert\psi_x(\mathbf{x},t)\rvert^2+\delta\right)\,
  \psi_x(\mathbf{x},t).
\end{align}
This is the set of equations governing the dynamics of exciton-polaritons that we numerically solve.

\subsection{Spin effects}

Important feature of exciton-polaritons is their spin, which allows to investigate their applications in spintronics~\cite{Sanvitto2012}.  In order to include spin in Eq.~(\ref{eq:gp_matrix2}), excitonic and photonic wave functions $\psi_{x,c}$ have to be computed separately for spin $\sigma=+1$ and $\sigma=-1$.  The coupling
constant $g$ is now replaced with two constants, $g_1$---quantifying coupling between excitons of the same spin and $g_2$---coupling between excitons of different spin.  This leads to following matrix equation
\begin{multline}
  \label{eq:gp_matrix2spin}
  i\hbar\,\dfrac{d}{dt}\begin{pmatrix}
    \psi_{c,-1}(\mathbf{x},t)\\
    \psi_{x,-1}(\mathbf{x},t)\\
    \psi_{c,+1}(\mathbf{x},t)\\
    \psi_{x,+1}(\mathbf{x},t)
  \end{pmatrix}
  =
  \begin{pmatrix}
    F_{-1}(\mathbf{x},t)\\
    0\\
    F_{+1}(\mathbf{x},t)\\
    0
  \end{pmatrix}
  +{}\\
  {}+\left[
    h^0+\begin{pmatrix}
      -i\hbar\,\frac{\gamma_c}{2}& 0& 0& 0\\
      0& -i\hbar\,\frac{\gamma_x}{2}
      +g_1\,\lvert\psi_{x,-1}(\mathbf{x},t)\rvert^2& 0& g_2\,\lvert\psi_{x,+1}(\mathbf{x},t)\rvert^2\\
      0& 0& -i\hbar\,\frac{\gamma_c}{2}& 0\\
      0& g_2\,\lvert\psi_{x,-1}(\mathbf{x},t)\rvert^2& 0& -i\hbar\,\frac{\gamma_x}{2} +g_1\,\lvert\psi_{x,+1}(\mathbf{x},t)\rvert^2
    \end{pmatrix}
  \right]\times{}\\
  {}\times\begin{pmatrix}
    \psi_{c,-1}(\mathbf{x},t)\\
    \psi_{x,-1}(\mathbf{x},t)\\
    \psi_{c,+1}(\mathbf{x},t)\\
    \psi_{x,+1}(\mathbf{x},t)
  \end{pmatrix},
\end{multline}
where $-1$ and $+1$ denote the spin $\sigma$ and
\begin{align}  F_{+1}(\mathbf{x},t)={}&F_{p_{+1}}\,e^{i\,(\mathbf{k}_{p_{+1}}\cdot\mathbf{x}-\delta_{\omega_{+1}}\,t)}\,e^{-\frac{\mathbf{x}^2}{2w_{x_{+1}}^2}},\\ F_{-1}(\mathbf{x},t)={}&F_{p_{-1}}\,e^{i\,(\mathbf{k}_{p_{-1}}\cdot\mathbf{x}-\delta_{\omega_{-1}}\,t)}\,e^{-\frac{\mathbf{x}^2}{2w_{x_{-1}}^2}},\\
  h^0={}&\begin{pmatrix}
    -\tfrac{\hbar^2\nabla^2}{2m_c}& \Omega_R& 0& 0\\
    \Omega_R& \delta& 0& 0\\
    0& 0& -\tfrac{\hbar^2\nabla^2}{2m_c}& \Omega_R\\
    0& 0& \Omega_R& \delta
  \end{pmatrix}.
\end{align}
The set of differential equations resulting from (\ref{eq:gp_matrix2spin}) obtains the following form
\begin{align}
  \label{eq:gp_linear_psic1spinM}
  i\hbar\,\frac{d}{dt}\,\psi_{c,-1}(\mathbf{x},t)={}&
  F_{-1}(\mathbf{x}, t)+\Omega_R\,\psi_{x,-1}(\mathbf{x},t)
  +\left(-i\hbar\,\frac{\gamma_c}{2}-\frac{\hbar^2\nabla^2}{2m_c}\right)\,
  \psi_{c,-1}(\mathbf{x},t),\\
  \label{eq:gp_linear_psic1spinP}
  i\hbar\,\frac{d}{dt}\,\psi_{c,+1}(\mathbf{x},t)={}&
  F_{+1}(\mathbf{x}, t)+\Omega_R\,\psi_{x,+1}(\mathbf{x},t)
  +\left(-i\hbar\,\frac{\gamma_c}{2}-\frac{\hbar^2\nabla^2}{2m_c}\right)\,
  \psi_{c,+1}(\mathbf{x},t),\\
  \label{eq:gp_linear_psix1spinM}
  i\hbar\,\frac{d}{dt}\,\psi_{x,-1}(\mathbf{x},t)={}&
  \Omega_R\,\psi_{c,-1}(\mathbf{x},t)  +\left(-i\hbar\,\frac{\gamma_x}{2}+g_1\,\lvert\psi_{x,-1}(\mathbf{x},t)\rvert^2+g_2\,\lvert\psi_{x,+1}(\mathbf{x},t)\rvert^2+\delta\right)\,
  \psi_{x,-1}(\mathbf{x},t),\\
  \label{eq:gp_linear_psix1spinP}
  i\hbar\,\frac{d}{dt}\,\psi_{x,+1}(\mathbf{x},t)={}&
  \Omega_R\,\psi_{c,+1}(\mathbf{x},t)  +\left(-i\hbar\,\frac{\gamma_x}{2}+g_1\,\lvert\psi_{x,+1}(\mathbf{x},t)\rvert^2+g_2\,\lvert\psi_{x,-1}(\mathbf{x},t)\rvert^2+\delta\right)\,
  \psi_{x,+1}(\mathbf{x},t).
\end{align}
This is the second set of polaritonic equations that we solve numerically using the EPCGP suite.

\subsection{Boundary conditions}
\label{ssec:boundary}

In order to solve the set of differential equations (\ref{eq:gp_linear_psic1})--(\ref{eq:gp_linear_psix1}) and (\ref{eq:gp_linear_psic1spinM})--(\ref{eq:gp_linear_psix1spinP}), it is necessary to set boundary conditions.  For initial time $t=0$ no quasiparticles are present in the microcavity thus, all wave functions are equal zero.  Additionally, $\psi_{c,x}$ vanish at the boundaries of the cavity.  This leads to the following set of conditions used in computations
\begin{align*}
  \psi_c(\mathbf{x},t=0)={}&0,\\
  \psi_x(\mathbf{x},t=0)={}&0,\\
  \psi_c(\lVert\mathbf{x}\lVert\geq L/2,t)={}&0,\\
  \psi_x(\lVert\mathbf{x}\lVert\geq L/2,t)={}&0,
\end{align*}
where $L$ denotes the radius of the microcavity and $\mathbf{x}_0$ lies in the center of the mesh.  Similar boundary conditions apply for the set of equations with spin.

\subsection{Parameters of the exciton-polariton equations}

Since the typical size of microcavities is of the order of micrometers and the lifetime of exciton-polaritons does not exceed $0.1\kern.25em\mathrm{ns}$, the most common units encountered in the literature in the description of exciton-polariton superfluids~\cite{Kavokin2007} are micrometers ($\mathrm{\mu m}$) and picoseconds ($\mathrm{ps}$).  They are complemented with a convenient unit of energy -- millielectronvolt ($\mathrm{meV}$).  This allows to express the wave functions in $\mathrm{\mu m}^{-1/2}$ for 1D condensate ($\mathrm{\mu m}^{-1}$ in the 2D case), frequency in $\mathrm{meV}$, decay rates in $\mathrm{ps}^{-1}$ and the interaction coefficient $g$ in $\mathrm{meV}\cdot\mathrm{\mu m}$($\mathrm{meV}\cdot\mathrm{\mu m}^2$ in the 2D case).  Pumping laser field $F_p$ is given in $\mathrm{meV}\cdot\mathrm{\mu m}^{-1/2}$ ($\mathrm{meV}\cdot\mathrm{\mu m}^{-1}$ in the 2D case) with momentum $k_p$ in $\mathrm{\mu m}^{-1}$ and detuning $\delta$ in $\mathrm{ps}^{-1}$.  Finally, physical constants expressed with these units equal to: reduced Planck constant $\hbar=0.6582\kern.25em\mathrm{meV}\cdot\mathrm{ps}$ and mass of afree electron $m_e=5.677\times10^3\kern.25em\mathrm{meV}\cdot\mathrm{\mu  m}^{-2}\cdot\mathrm{ps}^2$.

The typical values of the parameters of equations (\ref{eq:gp_matrix2}) and (\ref{eq:gp_matrix2spin}) are gathered in Table~\ref{tab:parameters}.  In case of 1D and 2D systems, the interaction coefficient may be converted with the following formula
\begin{equation}
  g^{\text{1D}}=\frac{g^{\text{2D}}}{\sqrt{2\pi d^2}},
\end{equation}
where $d$ is a width of a 1D microcavity.

\begin{table}[ht]
\begin{align*}
  \hbar={}&0.6582\,[\mathrm{meV}\cdot\mathrm{ps}]&&\text{-- reduced Planck constant,}\\
  \psi_c(x,t)={}&[1/\mu\mathrm{m}]&&\text{-- wave function for photons,}\\
  \psi_x(x,t)={}&[1/\mu\mathrm{m}]&&\text{-- wave function for polaritons,}\\
  F_p={}&[\mathrm{meV}/(\mu\mathrm{m})]&&\text{-- amplitude of the pump field,}\\
  k_p={}&[1/(\mu\mathrm{m})]&&\text{-- momentum of the pump field,}\\
  \delta_{\omega}={}&[1/(\mathrm{ps})]&&\text{-- detuning of the pump field,}\\
  w_x={}&[\mu\mathrm{m}]&&\text{-- spread of the pump,}\\
  g={}&0.01\,[\mathrm{meV}\cdot(\mu\mathrm{m})^2]&&\text{-- interaction coefficient,}\\
  \Omega_R={}&4.4\,[\mathrm{meV}]&&\text{-- Rabi frequancy,}\\
  \gamma_x={}&0.01\,[1/(\mathrm{ps})]&&\text{-- decay rate of an exciton,}\\
  \gamma_c={}&0.1\,[1/(\mathrm{ps})]&&\text{-- decay rate of a photon,}\\
  m_0={}&5.677\times 10^{3}\,[\mathrm{meV}/(\mu\mathrm{m}/(\mathrm{ps}))^2]&&\text{-- mass of a free electron},\\
  m_c={}&m_0\cdot 2\times 10^{-5}\,[\mathrm{meV}/(\mu\mathrm{m}/(\mathrm{ps}))^2]&&\text{-- effective mass of polaritons.}
\end{align*}
\caption{The parameters of the exciton-polariton equations (\ref{eq:gp_matrix2}) and (\ref{eq:gp_matrix2spin}) in the 2D case.}
\label{tab:parameters}
\end{table}

\section{Numerical methods}
\label{sec:numerics}

\subsection{Equations governing the dynamics of exciton-polaritons}

The two sets of equations presented in Section~\ref{sec:theory}, Eqs.~(\ref{eq:gp_linear_psic1})--(\ref{eq:gp_linear_psix1}) and (\ref{eq:gp_linear_psic1spinM})--(\ref{eq:gp_linear_psix1spinP}), share a similar structure.  Left-hand side of these equations is the first derivative of the individual wave function with respect to time.   The form of the right-hand side depends on the computed wave function.  In case of $\psi_c$ (as well as $\psi_{c,\sigma=\pm1}$) the terms include functions of time, linear functions of $\psi_x$ and $\psi_c$ and second order partial derivatives of the computed wave function $\psi_c$.  Equations defining $\psi_x$ ($\psi_{x,\sigma=\pm1}$) depend on the linear function of $\psi_c$ and both linear and nonlinear expressions involving $\psi_x$.  The main problem in solving these equations lies in a unique combination of complex-valued terms, nonlinearities and second order partial derivatives at their right-hand sides. Implicit (backward) methods, although usually stable, require solving algebraic equations, which make them unusable in the case of exciton-polariton equations.  In case of explicit (forward) methods it is more difficult to keep errors negligible.  Taking into account that our goal is to observe detailed evolution of the system with a finite time step, we compared the most important numerical methods of solving nonlinear differential equations, which could find appplication in computing the evolution of an exciton-polariton superfluid.

The most basic method is the Euler one.  It is simple and fast, but produces inaccurate results---an error introduced in the single step is of the order of $O(h^2)$, where $h$ is the step length.  The Runge--Kutta methods require additional stages of computation and therefore are slower, but much more accurate---the errors are of the order of $O(h^{n+1})$, where $n$ is the order of the method.  The methods based on the Richardson extrapolation (e.g.\ Bulirsch--Stoer algorithm) or predictor-corector algorithms are not suited to this task due to strong nonlinearities occurring in the exciton-polariton condensates.  They cause huge errors which must be compensated by small step size and computation time.  Comparison of the above explicit methods is shown in Table~\ref{tab:ode_methods}.  The best performance for a given precision was achieved by the adaptive Runge--Kutta algorithm.  However, the standard Runga--Kutta method of 4th order performed similarly well (only $25\%$ slower compared to the adaptive method) and its advantage lies in the constant step size, which makes solving of the evolution of the exciton-polaritons easier.  The other algorithms, although led to the same results, required more computing time.  Especially, advanced Bulirsch--Stoer and predictor-corector methods occurred to be slower than a relatively simple Runge--Kutta algorithm.

\begin{table}
\begin{tabular}{lccc}
\textit{Method}& \textit{Number of steps}& \textit{Relative error}& \textit{Computation time}\\\hline
Euler& $1.0\times10^6$& $10^{-5}$& $120\kern.25em\mathrm{s}$\\
Runge--Kutta (2nd order) & $2.0\times10^5$& $10^{-5}$& $50\kern.25em\mathrm{s}$\\
Runge--Kutta (4nd order) & $2.0\times10^3$& $10^{-5}$& $10\kern.25em\mathrm{s}$\\
Adaptive Runge--Kutta& $1.5\times10^3$& $10^{-5}$& $8\kern.25em\mathrm{s}$\\
Bulirsch--Stoer & $1.2\times10^5$& $10^{-5}$& $60\kern.25em\mathrm{s}$\\
Predictor-corector & $2.5\times10^5$& $10^{-5}$& $80\kern.25em\mathrm{s}$\\
\end{tabular}
\caption{The comparison of results of solving GPE for exciton-polariton superfluids with different numerical methods.  The test was based on a simulation of $1\kern.25em\mathrm{ps}$ evolution of a 1D condensate without spin.  All the programs were required to achieve relative accuracy of computations equal to $10^{-5}$.  The parameters used for computations are: $d=5$, $F_p=0.5$, $k_p=0$, $\delta=0$, $\delta_{\omega}=0$, $w_x=10$, $g=0.1$, $\Omega_R=4.4$, $\gamma_x=0.01$, $\gamma_c=0.1$.  The cavity size was set to $100\kern.25em\mathrm{\mu m}$ with $N=1000$ mesh nodes.}
\label{tab:ode_methods}
\end{table}

\subsection{The Runge--Kutta algorithm}
\label{ssec:RK}

The Runge--Kutta (RK) algorithm of the 4th order belongs to the family of the RK methods.  This approach evolved from the Euler method, where a differential equation $\tfrac{d}{dt} y(t) = f(t, y)$ is solved by substituting $\tfrac{d}{dt} y(t) \approx [y(t + h) - y(t)]/h$, where $h$ is a time step.  This leads to an approximation $y(t + h)\approx y(t) + hf(t, y)$.  In the Euler method the smaller the time step $h$ is, the more accurate is the solution but, at the same time, the computer program is more time-consuming and prone to errors resulting from finite-precision mathematical operations.  When $h>1$, the Euler method becomes unstable.

The 4th-order RK method stems from the Euler algorithm but introduces additional steps which improve accuracy and stability of computation.  These steps are denoted $k_1,\ldots,k_4$ and are computed in the following way
\begin{align}
  k_1={}& h\,f(y, t),\\
  k_2={}& h\,f(y+\tfrac{1}{2}\,k_1, t+\tfrac{1}{2}\,h),\\
  k_3={}& h\,f(y+\tfrac{1}{2}\,k_2, t+\tfrac{1}{2}\,h),\\
  k_4={}& h\,f(y+k_3, t+h).
\end{align}
Then,
\begin{equation}
  y(t+h)= y(t)+\tfrac{1}{6}\,(k_1+2\,k_2+2\,k_3+k_4).
\end{equation}
Within this approach, $k_1$ corresponds to the Euler method, $k_2$ and $k_3$ keep the corrections computed at the half-time step $t+\tfrac{1}{2}h$ and $k_4$ is the final correction calculated for the full step.  This method requires four computations of right-hand side of the equation, but due to better accuracy allows to use larger time steps and therefore performs better than the Euler algorithm.

In order to apply the RK method to Eqs.~(\ref{eq:gp_linear_psic1})--(\ref{eq:gp_linear_psix1}), both complex wave functions $\psi_{x,c}$ must be computed parallely in every step of the algorithm.  This is expressed in the following sequence of computations
\begin{align*}
  k_1^c={}& h\,f^c(\psi_x, \psi_c, t),\\
  k_1^x={}& h\,f^x(\psi_x, \psi_c, t),\\
  k_2^c={}& h\,f^c(\psi_x+\tfrac{1}{2}\,k^x_1, \psi_c+\tfrac{1}{2}\,k^c_1, t+\tfrac{1}{2}\,h),\\
  k_2^x={}& h\,f^x(\psi_x+\tfrac{1}{2}\,k^x_1, \psi_c+\tfrac{1}{2}\,k^c_1, t+\tfrac{1}{2}\,h),\\
  k_3^c={}& h\,f^c(\psi_x+\tfrac{1}{2}\,k^x_2, \psi_c+\tfrac{1}{2}\,k^c_2, t+\tfrac{1}{2}\,h),\\
  k_3^x={}& h\,f^x(\psi_x+\tfrac{1}{2}\,k^x_2, \psi_c+\tfrac{1}{2}\,k^c_2, t+\tfrac{1}{2}\,h),\\
  k_4^c={}& h\,f^c(\psi_x+k^c_3, \psi_c+k^c_3, t+h),\\
  k_4^x={}& h\,f^x(\psi_x+k^x_3, \psi_c+k^c_3, t+h),\\
  \psi_c(t+h)={}& \psi_c(t)+\tfrac{1}{6}\,(k_1^c+2\,k_2^c+2\,k_3^c+k_4^c),\\
  \psi_x(t+h)={}& \psi_x(t)+\tfrac{1}{6}\,(k_1^x+2\,k_2^x+2\,k_3^x+k_4^x),
\end{align*}
where $f^c(\psi_x, \psi_c, t)$ represents the right-hand side of Eq.~(\ref{eq:gp_linear_psic1}) and $f^x(\psi_x, \psi_c, t)$---of Eq.~(\ref{eq:gp_linear_psix1}).  Similarly, RK method applied to Eqs.~(\ref{eq:gp_linear_psic1spinM})--(\ref{eq:gp_linear_psix1spinP}) requires parallel computation of four sets of corrections ($k_n^{c,-1}$, $k_n^{c,+1}$, $k_n^{x,-1}$ and $k_n^{x,+1}$) related to four wave functions.  This makes the algoritm twice as long as in the case of spinless equations.

In order to represent the wave functions $\psi_{x,c}$ in a limited computer memory, the X-Y plane must be discretized in order to introduce a finite set of space coordinates.  Taking into account the form of equations describing exciton-polariton superfluid and the boundary conditions listed in Subsection~\ref{ssec:boundary}, a natural choice is to use a uniform mesh of $N$ nodes in a 1D case and a square mesh of $N\times N$ nodes for 2D superfluid.  The mesh should be centered in $\mathbf{x}=0$ and the distance between the consecutive nodes should be $\Delta_x$ and $\Delta_y$.  Hence, the effective size of the microcavity equals to $N\Delta_x$ for 1D system and $N\Delta_x\times N\Delta_y$ for a 2D microcavity.

Let us now focus on the right-hand side of the Eq.~(\ref{eq:gp_linear_psic1}).  It contains the nabla operator acting on the photonic wave function, $\nabla^2\psi_c(\mathbf{x},t)$.  In case of 1D condensate, this translates into $\nabla^2\psi_c(\mathbf{x},t)=\tfrac{d^2}{dx^2}\psi_c(x,t)$ which can be approximated for a uniform mesh by the following central finite difference formula
\begin{equation}
  \dfrac{d^2}{dx^2}\psi_c(x,t) = \dfrac{\psi_c(x-\Delta_x,t)-2\psi_c(x,t)+\psi_c(x+\Delta_x,t)}{\Delta_x^2},
\end{equation}
where $\Delta_x$ is the distance between mesh nodes.  Similarly, in two dimensions,
\begin{align}
  \nabla^2\psi_c(x,y,t)
  ={}& \left(\dfrac{d^2}{dx^2}+\dfrac{d^2}{dy^2}\right)\psi_c(x,y,t)\nonumber\\
  ={}& \dfrac{\psi_c(x-\Delta_x,y,t)-2\psi_c(x,y,t)+\psi_c(x+\Delta_x,y,t)}{\Delta_x^2}\nonumber\\
  &\qquad{}+\dfrac{\psi_c(x,y-\Delta_y,t)-2\psi_c(x,y,t)+\psi_c(x,y+\Delta_y,t)}{\Delta_y^2},
\end{align}
where $\Delta_x$ and $\Delta_y$ denote the distances between mesh nodes in the X and Y directions, respectively.

The total error accumulated in a single time step of the 4th order RK method is of the order of $O(h^5)$~\cite{NR}.  Errors resulting from discretization of the space are of the order of $O(\Delta_x^2)$ (1D system) and $O(\Delta_x^2\Delta_y^2)$ (2D system).  Summarizing, the smaller time step and the distance between mesh nodes, the more accurate result is obtained.

Stability of the presented Runge--Kutta method is imposed by the Courant--Friedrichs--Lewy (CFL) condition, which sets relation between the time step and mesh density.  In case of the 1D system the CFL condition amounts to
\begin{equation}
  \dfrac{\hbar}{m_c}\dfrac{h}{\Delta^2_x} \leq 1
\end{equation}
and for 2D one to
\begin{equation}
  \dfrac{\hbar}{m_c}\dfrac{h}{\Delta^2_x} + \dfrac{\hbar}{m_c}\dfrac{h}{\Delta^2_y} \leq 1.
\end{equation}

\subsection{Data structures}

The wave functions, $\psi_{x,c}$ take complex values, represented in the computer memory by a pair of double-precision IEE754 floating-point numbers.  The discrete mesh is stored as a 1D or 2D array of values of the nodes.  In this way, a single memory allocation of $16N$ bytes in 1D case ($16N^2$ in 2D case) is sufficient to store the results.

Data structures used by the 4th order RK numerical integration method are also 1D and 2D arrays of complex numbers.  Since the algorithm requires computing coefficients $k_1,\ldots,k_4$ for each mesh node, the total memory requirement is $16\cdot 6N$ bytes for 1D mesh ($16\cdot 6N^2$ for 2D mesh).  In case of typical value of $N=1000$ the amount of memory required by the program to run is of the order of several hundred megabytes.

\subsection{Optimization}

The computational complexity of the RK algorithm presented in the Subsection~\ref{ssec:RK} is of the order of $O(N\cdot M)$, where $N$ is the number of mesh nodes and $M$ is the number of computed time steps.  In case of 2D superfluid, this value grows to $O(N^2\cdot M)$.

The main advantage of EPCGP suite lies in the specially developed optimization of the RK method used for solving the equations describing exciton-polariton superfluid.  Up to date, most optimization methods have been based on dividing the mesh into subsets (tiles) of equal sizes and performing the RK steps for all these subsets in parallel.  Synchronization points lie between algorithm steps and at each iteration of the main loop of the program. This allows to reduce the required time $T$ times, where $T$ is the number of computing cores involved.

Our method takes into account the specific form of the right-hand-sides of the equations describing exciton-polaritons.  We noticed that the terms may be put in groups of different properties: (a) expressions not depending on $\psi_{c,x}$, e.g.\ $F(\mathbf{x},t)$, (b) terms proportional to the computed wave functions, e.g.\ $\Omega_R\psi_{c,x}$, (c) terms proportional to the square of the functions, and (d) terms proportional to $\lvert\psi_x\rvert^2 \psi_x$.  Therefore, the EPCGP performs the individual steps of the RK algorithm, denoted as $k_1,\ldots,k_n$, in the order (a)--(d).  In each substep, the corresponding power of $\psi_{x,c}$ is computed and stored for the next substep.  Additionally, the whole set of terms for all mesh nodes is computed in a single step with the help of vector commands, available for the contemporary computing platforms.  In this way, when $N$ is sufficiently small, the computation time is reduced to $O(M)$ for both 1D and 2D systems.  In case of larger values of $N$, all operations are divided into sets of $64$--$512$ nodes, which are suitable for hardware vector operations.  This allows to set the computation time of the order of $O(N\cdot M / V)$ or $O(N^2\cdot M / V)$, where $V$ is the maximal target of the vector operation.

The EPCGP suite utilizes OpenMP parallel framework~\cite{OpenMP,OpenMP_book} as well as MPI (Message Passing Interface)~\cite{MPI,MPICH} in order to split the computational task.  This solution fits well to the typical environment of computer clusters. The main process/thread is responsible for gathering the data and saving results to disk files.

Both OpenMP and MPI technologies are optional and the suite gives the same results regardless they are used or not.  They can be turned on or off by setting compiler directives in the source code of the program.

\subsection{Discrete Fourier transform}

The algorithms involved in solving the equations describing the exciton-polariton superfluid in the EPCGP suite compute the evolution in the configuration space.  However, the suite has been additionally equipped with procedures calculating discrete Fourier transforms (DFTs) and the inverse DFTs of the wave functions $\psi_{x,c,\sigma=\pm1}$. The goal of including these procedures is to ease presentation of results in the frequency space, which is often encountered in the literature devoted to exciton-polaritons.

The DFT of a 1D wave function $\psi(x)$ over the mesh of $N$ nodes and distance $\Delta_x$ between the nodes where $x=x_0 + k\Delta_x$ is given by
\begin{align}
  \tilde{\psi}(k_x)
  = \dfrac{1}{\sqrt{N}}\sum_{k=0}^{N} \psi(x_0 + k\Delta_x) e^{-i[k_x\,(x_0 + k\Delta_x)]},
\end{align}
with inverse transform defined as
\begin{equation}
  \psi(x)
  = \dfrac{1}{\sqrt{N}}\sum_{k=0}^{N} \tilde{\psi}(k_{x_0} + k\Delta_{k_x}) e^{i[x\,(k_{x_0} + k\Delta_{k_x})]},
\end{equation}
where, under assumption that $x=0$ lies in the middle of the mesh,
\begin{align}
  k_{x_0}={}& -\dfrac{\pi}{\Delta_x},\\
  \Delta_{k_x} ={}& \dfrac{2\pi}{N\Delta_x}.
\end{align}
There are numerous known algorithms of computing DFTs.  The EPCGP suite utilizes the Cooley--Tukey Fast Fourier Transform (FFT) algorithm~\cite{Smith2007} in order to achieve the best possible performance.  The procedure is further optimized for use with OpenMP library.  The inverse DFT shares the same code due to the similarity of FFT and inverse FFT algorithms.  Two-dimensional case is computed with two passes of a 1D FFT procedure.

\section{Examples and Applications}
\label{sec:applications}

The EPCGP suite was utilized to obtain numerous results for different input parameters.  Both 1D and 2D exciton-polariton superfluids were investigated and spin effects were optionally taken into account.  Stability of results was ensured by the proper relation between mesh density and size of the time step, found with help of the Courant--Friedrichs--Lewy condition, presented in Subsection~\ref{ssec:RK}.

\begin{figure*}\centering
\includegraphics[width=10cm]{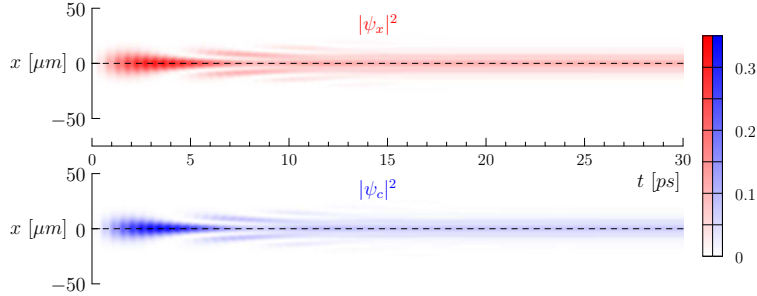}
\caption{Time evolution of 1D exciton-polariton superfluid without spin effects.  The red plot depicts the evolution of the polaritonic wave function $\lvert\psi_x\rvert$ whereas blue---photonic wave function $\lvert\psi_c\rvert$.  The computations were performed for a condensate of width $d=5\kern.25em\mu\text{m}$, pump amplitude $F_p=0.5\kern.25em\text{meV}\cdot\sqrt{\mu\text{m}}$, detuning of the pump $\delta_{\omega}=5\kern.25em 1/\text{ps}$, Rabi frequency $\Omega_R = 4.4\kern.25em\text{meV}$, interaction coefficient $g=-50\kern.25em\text{meV}\cdot(\mu\text{m})^2$ and decay rates $\gamma_c=0.5\kern.25em 1/\text{ps}$, $\gamma_x=0.05\kern.25em 1/\text{ps}$.  After $t=15\kern.25em\text{ps}$ the system arrives at a stationary solution.}
\label{fig:example1}
\end{figure*}

\begin{figure*}\centering
\includegraphics[width=10cm]{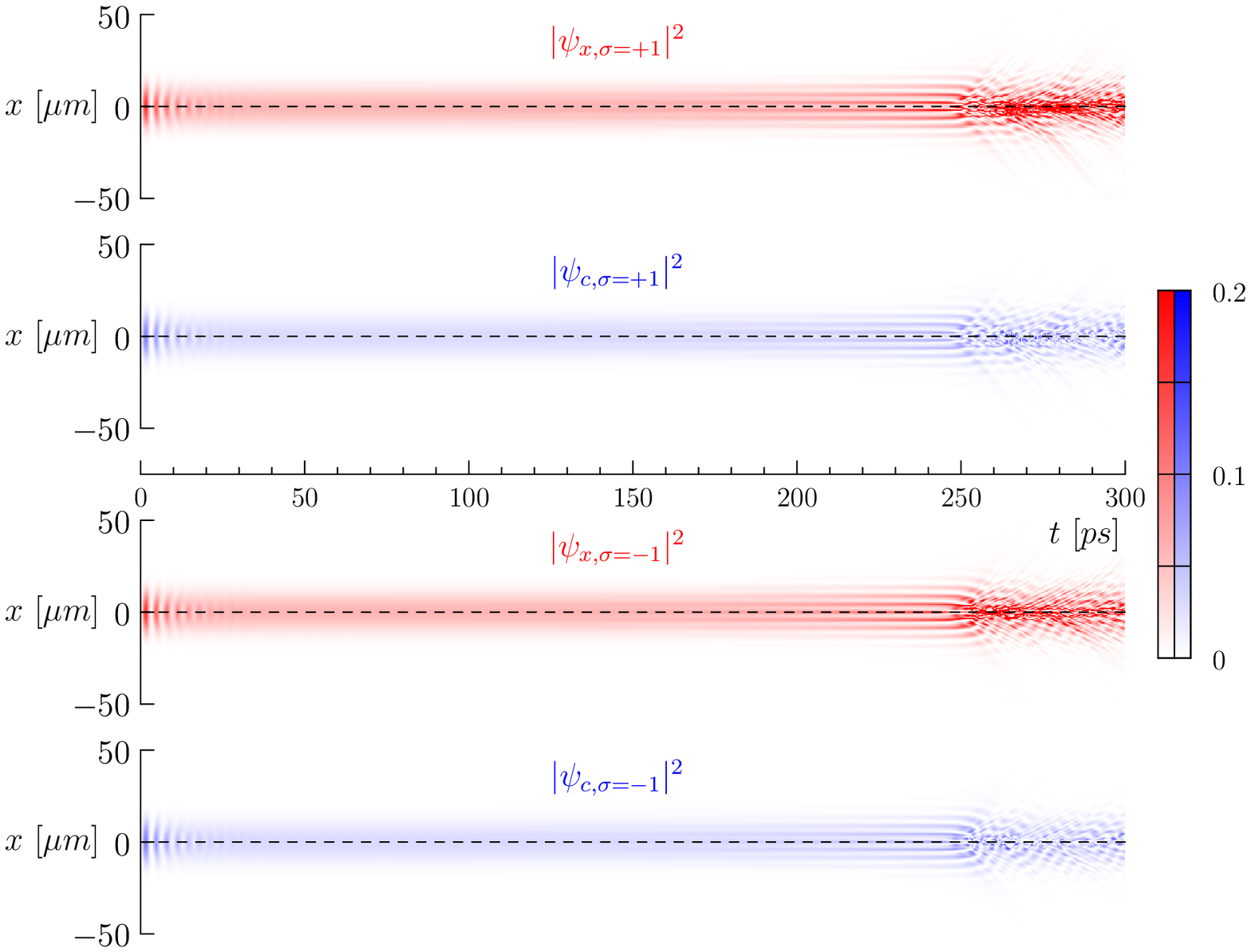}
\caption{Time evolution of 1D exciton-polariton superfluid with spin effects.  The red plots depict the evolution of the polaritonic wave functions $\lvert\psi_{x,\sigma=\pm1}\rvert$ whereas blue -- photonic wave functions $\lvert\psi_{x,\sigma=\pm1}\rvert$ for spins $\sigma\in\{-1,1\}$.  The computations were performed for a condensate of width $d=5\kern.25em\mu\text{m}$, pump amplitude $F_p=0.5\kern.25em\text{meV}\cdot\sqrt{\mu\text{m}}$, detuning of the pump $\delta_{\omega}=5\kern.25em 1/\text{ps}$, Rabi frequency $\Omega_R = 4.4\kern.25em\text{meV}$, interaction coefficients $g_1=50\kern.25em\text{meV}\cdot(\mu\text{m})^2$, $g_2=-10\kern.25em\text{meV}\cdot(\mu\text{m})^2$ and decay rates $\gamma_c=0.5\kern.25em 1/\text{ps}$, $\gamma_x=0.05\kern.25em 1/\text{ps}$. After $t=20\kern.25em\text{ps}$ the system arrives at a stationary solution, but long evolution times $t>250\kern.25em\text{ps}$ reveal instabilities.}
\label{fig:example2}
\end{figure*}

\begin{figure*}\centering
\includegraphics[width=10cm]{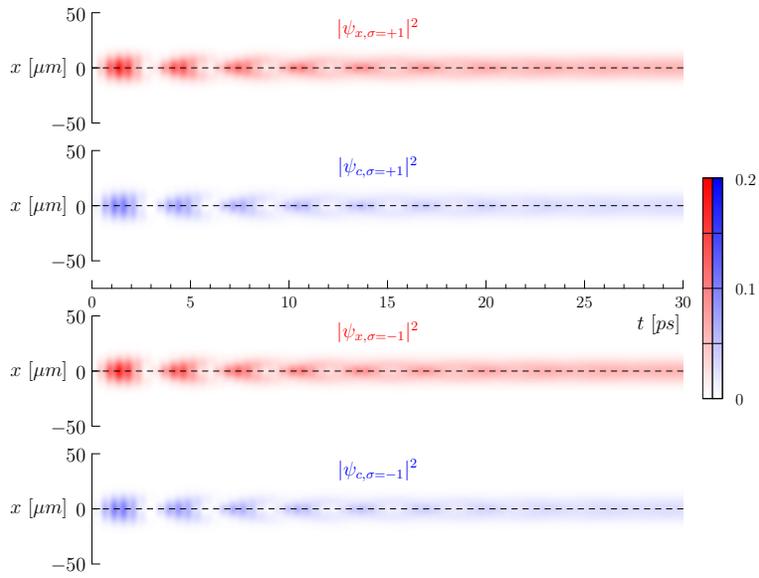}
\caption{Time evolution of 1-dimensional polaritonic BEC with spin effects, modeled with the set of G--P equations.  The red plots depict the evolution of the polaritonic wave functions $\lvert\psi_{x,\sigma=\pm1}\rvert$ whereas blue -- photonic wave functions $\lvert\psi_{x,\sigma=\pm1}\rvert$ for spins $\sigma\in\{-1,1\}$.  The computations were performed for a condensate of width $d=5\kern.25em\mu\text{m}$, pump amplitude $F_p=0.5\kern.25em\text{meV}\cdot\sqrt{\mu\text{m}}$, detuning of the pump $\delta_{\omega}=5\kern.25em 1/\text{ps}$, Rabi frequency $\Omega_R = 4.4\kern.25em\text{meV}$, interaction coefficients $g_1=50\kern.25em\text{meV}\cdot(\mu\text{m})^2$, $g_1=-2\kern.25em\text{meV}\cdot(\mu\text{m})^2$ and decay rates $\gamma_c=0.5\kern.25em 1/\text{ps}$, $\gamma_x=0.05\kern.25em 1/\text{ps}$. The solution reveals oscillations, which vanish for $t>20\kern.25em\text{ps}$.}
\label{fig:example3}
\end{figure*}

\begin{figure}[p]
  \begin{center}
    \raisebox{3.5cm}{a)}
    \includegraphics[height=3.5cm]{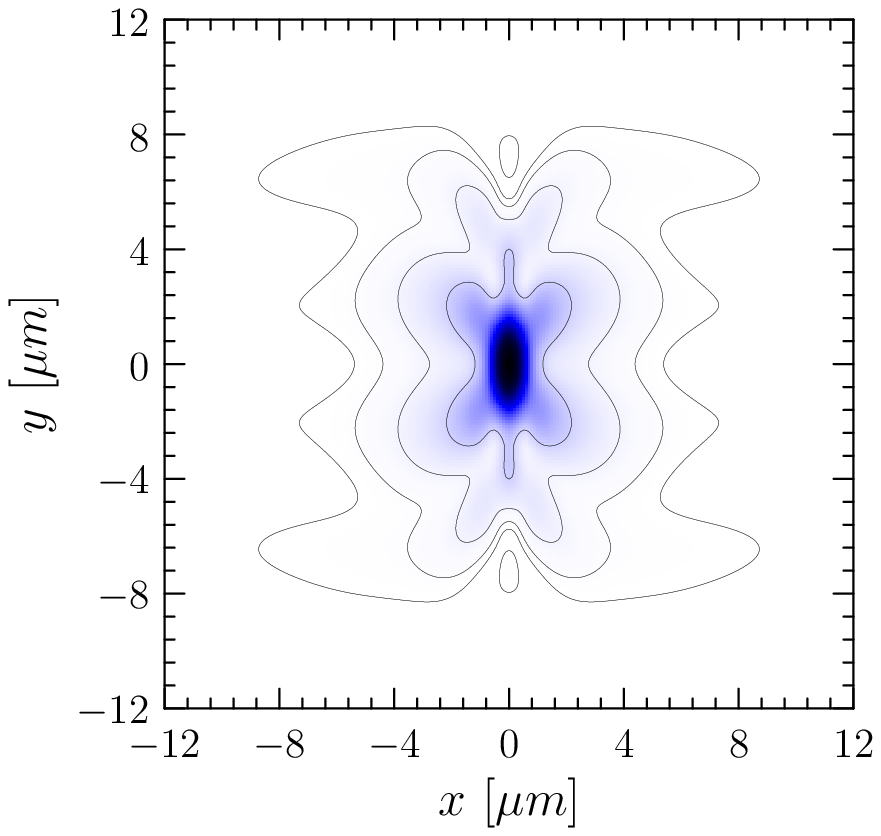}
    \includegraphics[height=3.5cm]{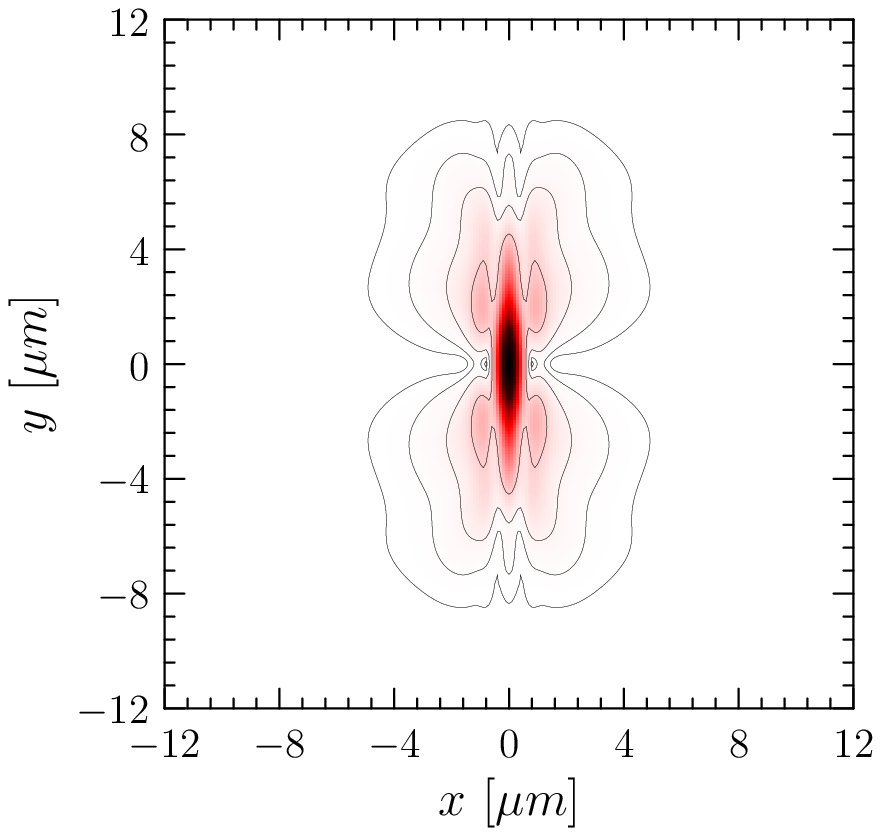}\\
    \raisebox{3.5cm}{b)}
    \includegraphics[height=3.5cm]{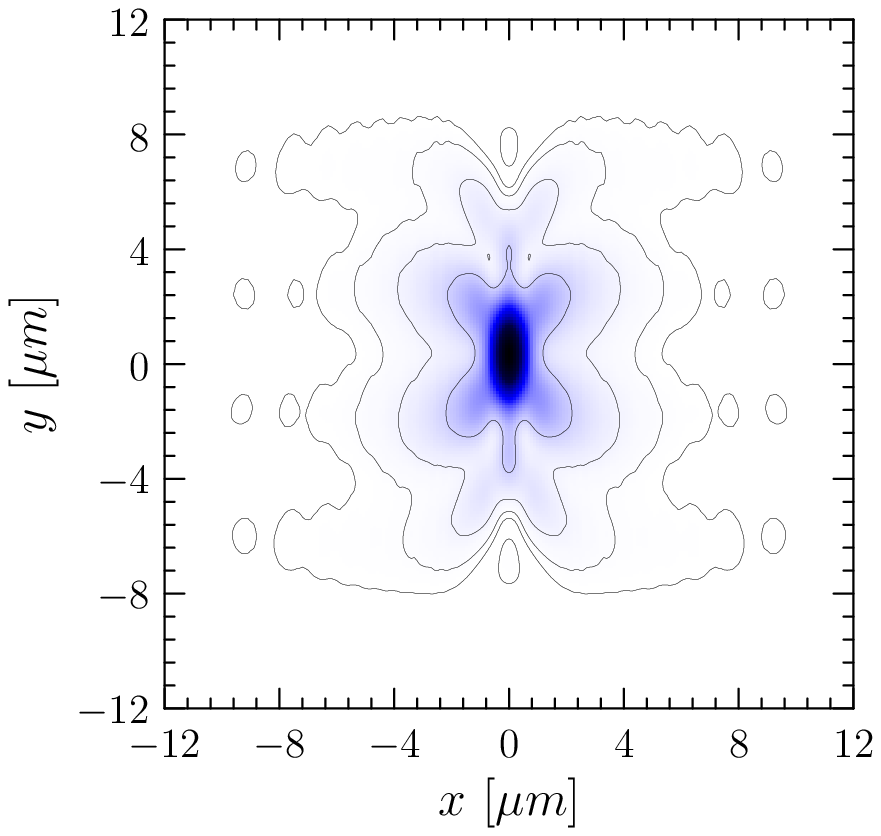}
    \includegraphics[height=3.5cm]{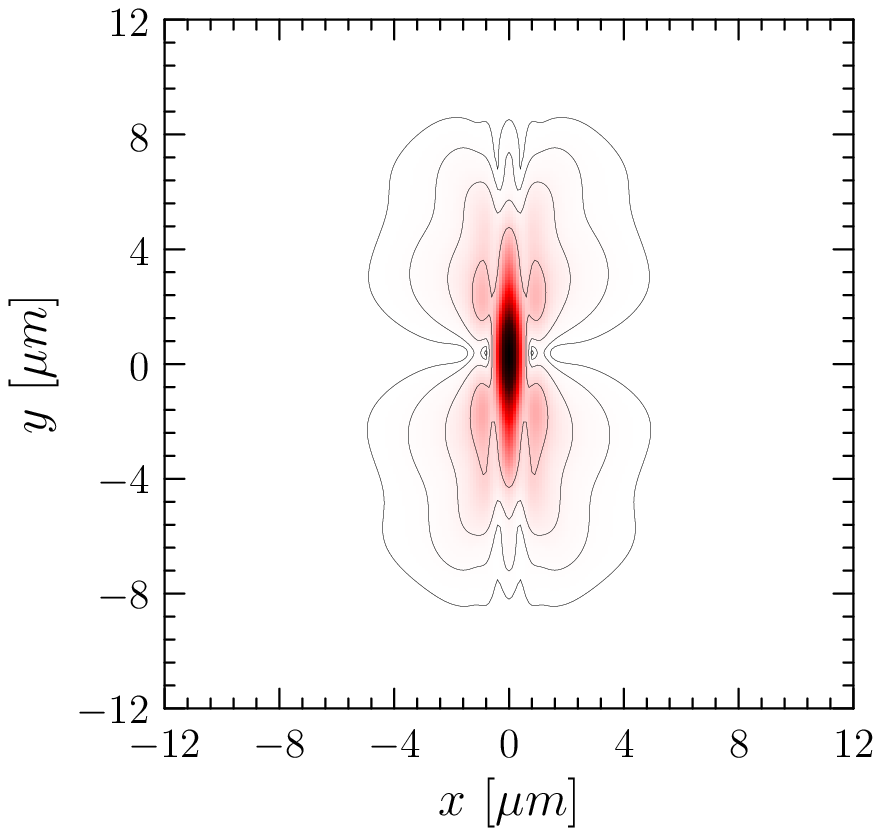}\\
    \raisebox{3.5cm}{c)}
    \includegraphics[height=3.5cm]{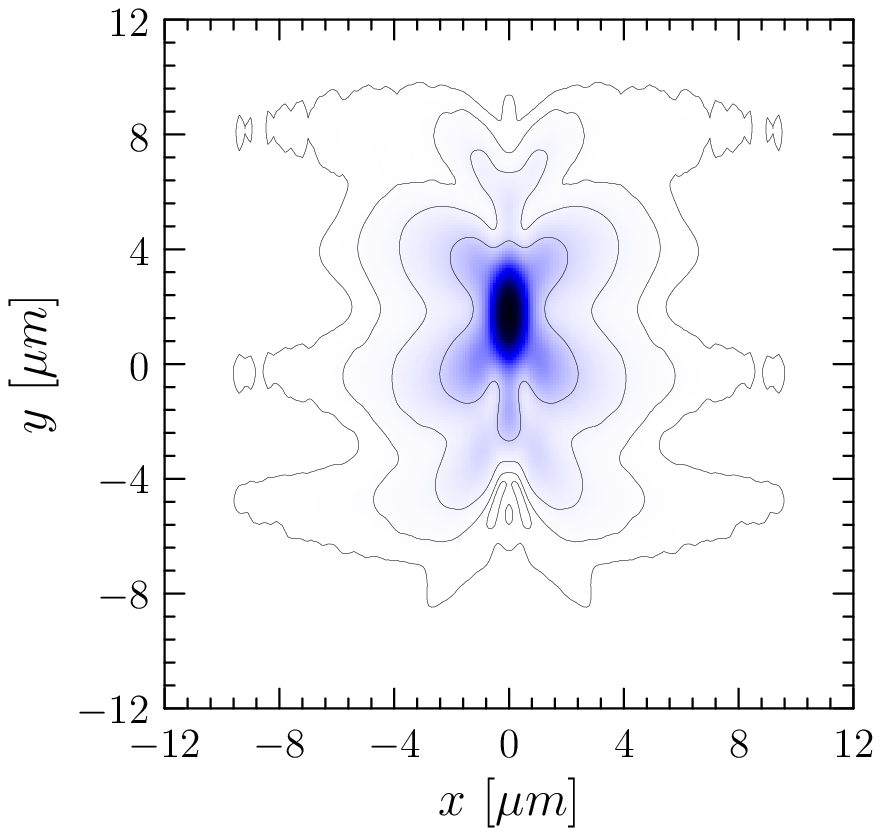}
    \includegraphics[height=3.5cm]{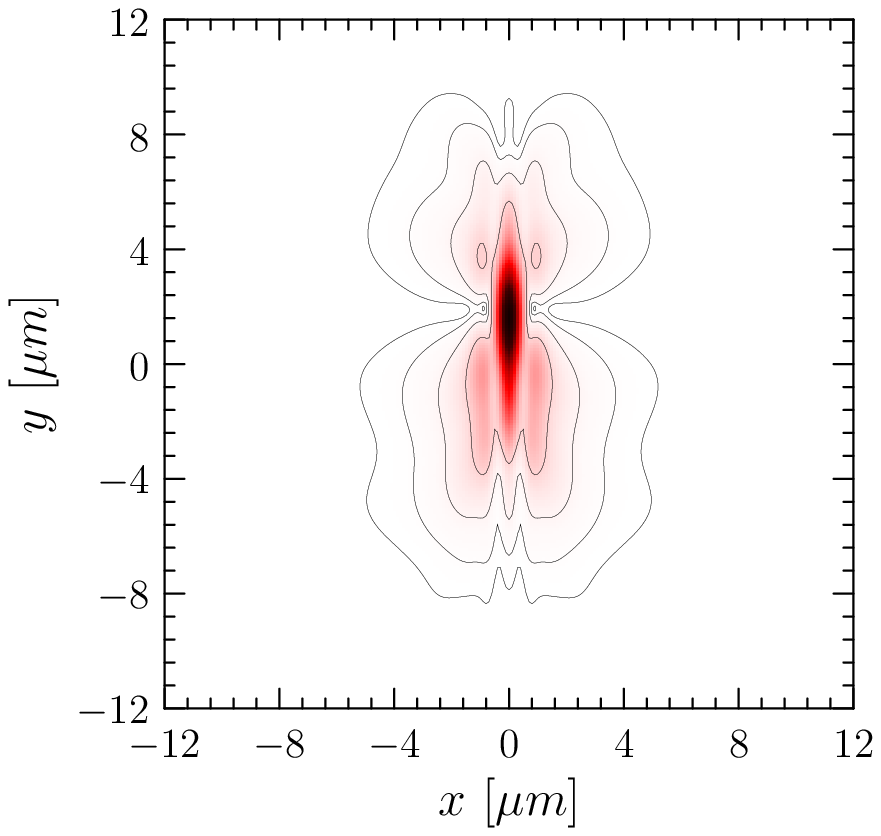}
  \end{center}
  \caption{The evolution of $\psi_c(x, y)$ (left column) and
    $\psi_x(x, y)$ (right column), being the solution of
    Eqs.~(\ref{eq:gp_linear_psic1})-(\ref{eq:gp_linear_psix1}) for a
    specially constructed initial state, computed for a) $t=0$, b)
    $t=1ps$, c) $t=5ps$.  The results were obtained for the following
    parameters: $F_p=0$, detuning of the pump $\delta_{\omega}=0$,
    Rabi frequency $\Omega_R = 1\kern.25em\text{meV}$, interaction
    coefficient $g=0$ and decay rates $\gamma_c=\gamma_x=0$.}
  \label{fig:example2D}
\end{figure}

Here we present several plots showing the evolution of a 1D condensate.  Fig.~\ref{fig:example1} presents typical simulation where spin is neglected.  Two wave functions are depicted as particle number densities $\lvert\psi_{x,c}\rvert^2$ changing in time.  The investigated system arrives at a stationary solution after $t=15\kern.25em\mathrm{ps}$. Fig.~\ref{fig:example2} presents simulation for similar parameters as for Fig.~\ref{fig:example1}, but this time spin is taken into account.  In this case, the condensate reveals instabilities after a long evolution time $t>250\kern.25em\text{ps}$. Fig.~\ref{fig:example3} displays similar case but for $g_1\not=g_2$.  The solution reveals oscillations, which vanish for $t>20\kern.25em\text{ps}$.  Finally, Fig.~\ref{fig:example2D} depicts an evolution of a 2D system for a given initial state.

\section{EPCGP Program Suite}
\label{sec:suite}

The EPCGP is a program suite developed for numerical modeling of evolution of an exciton-polariton superfluid in four different scenarios:
\begin{itemize}
\item 1-dimensional exciton-polariton superfluid, neglected spin effects,
\item 2-dimensional exciton-polariton superfluid, neglected spin effects,
\item 1-dimensional exciton-polariton superfluid, spin effects included,
\item 2-dimensional exciton-polariton superfluid, spin effects included.
\end{itemize}
This allows to perform numerical simulations in most of typical cases.

\subsection{Compilation of the Programs}

In order to compile and test EPCGP program suite for Linux/Unix operating systems with GNU utilities (GNU make, GNU compiler collection) it is enough to run the following commands in a directory containing unpacked source code
\begin{verbatim}
    make
    make check
\end{verbatim}
In case of Linux/Unix with GNU make and Intel C Compiler, one has to modify \verb|Makefile|, replacing \verb|gcc| with \verb|icc| and \verb|-fopenmp| option with \verb|-openmp|.

For other platforms and compilers: one should compile all the program files with a standard C/C++ compiler and OpenMP extensions turned on. If compiler does not offer OpenMP extensions, the program will still work but will not utilize multiple cores or processors.

\subsection{Description of the Program Suite}

Program computes time evolution of distribution of photons and polaritons in the polaritonic semiconductor microcavity.  The results are saved to a disk file in a format suitable for further processing, e.g.\ visualization.  At each step it also prints some indicators, used for validation of the computations, namely: norms of wave functions, current evolution time and the amount of time spent in computations.

Program parameters are set in the source code, by assigning values to the predefined variables.
\begin{verbatim}
// Parameters of the setup.
const double wkx = 4.0;                 // Spread of the initial Gaussian along k_x.
const double wky = 0.25;                // Spread of the initial Gaussian along k_y.
const double A = 0.5;           // Amplitude of the initial Gaussian.
const double m = 1.0;                   // Boson mass [meV].
const double me = 1.0;                  // Effective mass of polaritons [meV].
const double wb = 10;                   // The width of the “safe” window [µm].
const double cavsizex = 12;             // Cavity size along X axis [µm].
const double cavsizey = 12;             // Cavity size along Y axis [µm].
const double h = 0.0001;                // Time step [ps].

// Parameters of the simulation.
const int xsize = 121;                  // Number of mesh nodes along X axis.
const int ysize = 121;                  // Number of mesh nodes along Y axis.
const double sx = cavsizex / (xsize - 1); // Spatial step along X axis.
const double sy = cavsizey / (ysize - 1); // Spatial step along Y axis.
const double xmin = -(cavsizex / 2); // Lower boundary of X coordinate.
const double xmax = (cavsizex / 2); // Upper boundary of X coordinate.
const double ymin = -(cavsizey / 2); // Lower boundary of Y coordinate.
const double ymax = (cavsizey / 2); // Upper boundary of Y coordinate.
const double kxmin = -pi / sx;          // Lower boundary of k_x coordinate.
const double kxmax = pi / sx;           // Upper boundary of k_x coordinate.
const double kymin = -pi / sy;          // Lower boundary of k_y coordinate.
const double kymax = pi / sy;           // Upper boundary of k_y coordinate.
const double ksx = 2 * pi / cavsizex; // Frequency step along k_x axis.
const double ksy = 2 * pi / cavsizey; // Frequency step along k_y axis.
const int nsteps = 200000;                       // Number of simulation iterations.
const int print_tstep = 1000;                    // Print every nth time step.
const int print_xstep = 1;               // Print every nth spatial point along X axis.
const int print_ystep = 1;               // Print every nth spatial point along Y axis.
\end{verbatim}
After setting the parameters, program has to be recompiled in order to achieve best optimization of the machine code.  Running of the program is not interactive and therefore allows for execution in the environment of computer clusters.  In order to set the maximal number of cores or processors utilized by the program one could use \verb|OMP_NUM_THREADS| environment variable.  If not set, this value is obtained from the operating system and all computing cores are utilized.

\section{Conclusions}

In this paper we presented the EPCGP program suite developed for numerical modeling of an exciton-polariton superfluid in a semiconductor microcavity.  It solves sets of nonlinear differential equations describing exciton-polaritons  in one and two dimensions.  They are based on the Gross--Pitaevskii equation, but describe separate excitonic and photonic wave functions.  We listed and briefly described the parameters of the modeled system.

Numerical procedures included in the EPCGP suite utilize the Runge--Kutta method of the 4th order.  The choice of the numerical algorithm was preceded by the analysis of applicability and performance of several numerical routines in solving the exciton-polariton equations for typical values of experimental parameters.  We optimized the Runge--Kutta algorithm for this task by rearranging algebraic operations in order to avoid unnecessary intermediate steps.  Furthermore, the suite code utilizes parallel OpenMP and MPI compiler extensions as well as efficient vector operations built into modern processors. This makes the EPCGP suite a fast and convenient tool for reliable numerical modeling of exciton-polariton superfluids.  It may find applications in theoretical investigation of properties and phenomena observed in exciton-polariton systems.

\section*{Acknowledgments}

MS, OV, AB were supported by the EU 7FP Marie Curie Career Integration
Grant No. 322150 ``QCAT'', NCN grant No. 2012/04/M/ST2/00789, MNiSW
co-financed international project No. 2586/7.PR/2012/2 and MNiSW
Iuventus Plus project No. IP 2014 044873.

\end{document}